\documentclass[showpacs,amssymb,twocolumn,floatfix,prb,aps]{revtex4-1}

\usepackage{amssymb,amsfonts,bm}
\usepackage{graphics,graphicx}

\begin{document}

\title{Separation of Wigner structures for 2D equimolar binary mixtures of 
Coulomb particles}

\author{Igor Trav{\v e}nec}
\affiliation{Institute of Physics, Slovak Academy of Sciences, 
D\'ubravsk\'a cesta 9, 84511 Bratislava, Slovakia}
\author{Ladislav \v{S}amaj}
\affiliation{Institute of Physics, Slovak Academy of Sciences, 
D\'ubravsk\'a cesta 9, 84511 Bratislava, Slovakia}

\begin{abstract}
We study the lowest energy configurations of an equimolar binary mixture 
of classical pointlike particles with charges $Q_1$ and $Q_2$, such 
that $q=Q_2/Q_1\in [0,1]$.
The particles interact pairwisely via 3D Coulomb potential and are confined 
to a 2D plane with a homogeneous neutralizing background charge density. 
In a recent paper by M. Antlanger and G. Kahl 
[Cond. Mat. Phys. {\bf 16}, 43501 (2013)], 
using numerical computations based on evolutionary algorithm, 
six fully mixed structures were identified for $0\le q\lesssim 0.59$, while 
the separation of $Q_1$ and $Q_2$ pure hexagonal phases minimizes the energy 
for $0.59\lesssim q<1$. 
Here, we introduce a novel structure which consists in the separation of 
two phases, the pure hexagonal one formed by a fraction of particles with 
the larger charge $Q_1$ and the other mixed one containing different numbers 
of $Q_1$ and $Q_2$ charges. 
Using an analytic method based on an expansion of the interaction energy
in Misra functions we show that this novel structure provides the lowest 
energy in two intervals of $q$ values, $0<q\lesssim 0.04707$ and 
$0.58895\lesssim q\lesssim 0.61367$.
This fact might inspire numerical methods, for both Coulomb and Yukawa
interactions, to test more general separations which go beyond the separation
of two pure phases.
\end{abstract}


\date{\today}

\maketitle

\section{Introduction} \label{Intro}
Various binary systems of particles with pairwise interactions, confined 
to a plane, have been studied intensively.
They are known to exhibit a richer variety of ground state (zero-temperature)
crystal structures than the one-component ones.\cite{Bonsall}
2D binary like-charge crystals with Yukawa (screened Coulomb) interactions 
were analyzed in detail by using Monte-Carlo simulations,\cite{AML} 
a plenty of triangle, square and rhombic lattices were identified.
Such structures were found also experimentally,\cite{rev,Nunomura}
especially in colloidal suspensions.
The Coulomb potential is a limiting case of the Yukawa one, so
it is not surprising that similar structures appear in 2D binary
Coulomb systems as well, as was shown in numerical simulations \cite{Ant}
based on a combination of Ewald summation techniques\cite{Mazars11} and 
the evolutionary algorithm (EA).\cite{Gott}

Some binary systems tend to separate if the parameters (size, charge, \ldots)
of the two components are similar.
This was observed in the case of hard disks when their two diameters are 
comparable,\cite{LH,Schmidt} for the Yukawa binary systems \cite{AML} 
as well as for long-ranged Coulomb systems of like charges \cite{Ant} when 
the charges do not differ too much.
Analogous behavior, called demixing, was reported also for Lenard-Jones 
mixtures \cite{MPS} and binary mixtures with dipole interactions \cite{AMLo}.
The impact of the interaction type on the resulting ground-state structures 
is still far from being understood.

In this work, we study the lowest energy configurations of a binary mixture 
of classical pointlike particles with charges $Q_1$ and $Q_2$ of the same sign.
We introduce the parameter 
\begin{equation} \label{defq}
q\equiv \frac{Q_2}{Q_1} .
\end{equation} 
Due to the symmetry $q\to 1/q$, one can restrict itself to the interval 
$q\in [0,1]$, i.e. $Q_1$ is larger than $Q_2$.
We consider the equimolar case, i.e. the numbers of species $N_1=N_2=N/2$,
where $N$ is the total number of particles.
The particles, confined to a 2D plane with a neutralizing background charge
density, interact via the 3D Coulomb $1/r$ pairwise potential. 

Six fully mixed structures were identified for $0\le q\lesssim 0.59$, while 
the separation of pure $Q_1$ and $Q_2$ hexagonal phases minimizes the energy 
for $0.59\lesssim q<1$.\cite{Ant}   
Our main contribution to the field consists in the introduction of a novel 
structure which consists in the separation of two phases, the pure hexagonal 
one formed by a fraction of particles with the larger charge $Q_1$ and 
the other completely mixed one containing different numbers of 
$Q_1$ and $Q_2$ charges. 
Using an analytic method based on an expansion of the interaction energy
in Misra functions we show that this novel structure possesses the lowest 
energy in two intervals of $q$, $0<q\lesssim 0.04707$ and 
$0.58895\lesssim q\lesssim 0.61367$.

The paper is organized as follows.
In Sec. \ref{Sec2}, we introduce the structures found by numerical 
simulation.\cite{Ant}
There are six completely mixed structures, i. e. those with the same number 
of the two components in one periodic cell.
We calculate their energies per particle using analytic formulas, 
with an extremely high precision.
The energies are compared to the energy of the separation of two pure phases. 
In Sec. \ref{Sec3}, we introduce the novel structure, i.e. the separation
of a pure $Q_1$-phase and a specifically mixed $Q_1-Q_2$ phase and calculate 
its energy.
The results for the phase diagram are presented in Sec. \ref{Sec4}.
The impact of our study to numerical simulations and other possible applications
of the method are indicated in Conclusion.
Long formulas, with a sketch of their derivation, are presented in Appendices.

\section{Known ground-state structures} \label{Sec2}
The equimolar binary system of $N_1=N/2$ $Q_1$-charges and $N_2=N/2$ 
$Q_2$-charges is confined to a plane of surface $S$, the particle density 
being $\rho=N/S$. 
The particle charge density $\sigma$ is given by $S\sigma=(Q_1+Q_2)N/2$, i.e.
\begin{equation} \label{sigro}
\sigma=\rho\ Q_1 \frac{1 + q}{2}.
\end{equation}
The particle system is immersed in a neutralizing homogeneous background 
of the charge density $-\sigma$.
The Coulomb potential energy of charges $q$ and $q'$ at distance $r$ is
given by $q q'/r$. 

We express the energy of all structures using the method based on a
sequence of transformations for Coulomb lattice sums.\cite{Samaj}
In contrast to the energies of bilayer Wigner crystals, which are expressed
as series of generalized Misra functions, the energies of the present planar
Wigner structures require the introduction of the standard Misra 
functions\cite{Misra}, see Eq. (\ref{Misra}) of Appendix \ref{ApA}.
The particles can always be divided into infinite lattice subsets 
(with the corresponding decomposition of the neutralizing background) 
in such a way that the energy per particle is expressible by using two types 
of lattice summations $\Sigma_1$ and $\Sigma_2$, 
defined and expressed as series of Misra functions in Appendix \ref{ApB}.
These series are extremely quickly convergent and in our calculations we
take first five expansion terms.
The minimization of the energy with respect to lattice parameters
reproduces the exact ground-state energy up to 15 decimal digits;
this was checked on the well-known Madelung constant of the hexagonal
lattice and, for a few chosen structures, by including also the sixth 
and seventh expansion terms. 

Next we recall the six completely mixed structures, which have 
the same number of $Q_1$ and $Q_2$ particles per cell, found in 
EA numerical simulations.\cite{Ant}
Since some of the structures were described only partially, we provide
a detailed definition of the relevant structures in order to specify 
their energies with a high accuracy.
This is needed because energy differences among structures are very small,
especially in the region of small values of $q$.  
At the end, we write down the energy per particle for the separation
of two pure phases. 

\subsection{Structure I}

\begin{figure}[]
\begin{center}
\includegraphics[clip,width=0.34\textwidth]{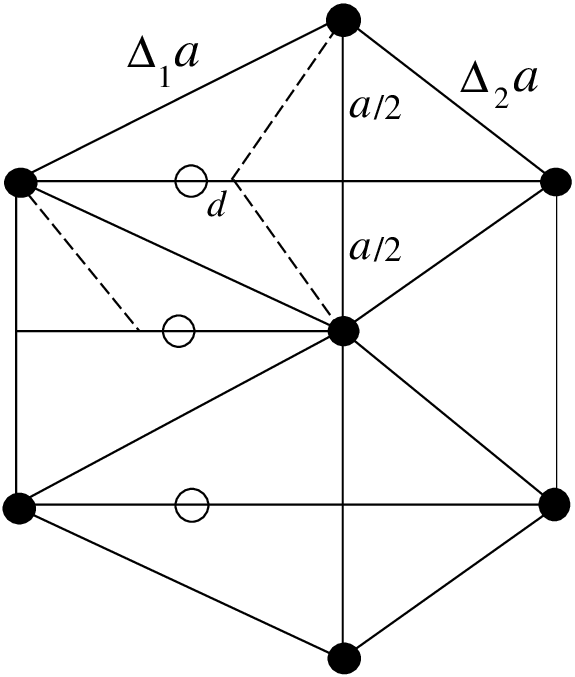}
\caption{Structure I. Full disks correspond to particles with charge $Q_1$, 
the empty ones to $Q_2$.  
Dashed lines in one of the left triangles meet at the circumcenter 
passing through three $Q_1$ particles.}
\label{fig:StructureI}
\end{center}
\end{figure}

For small values of $q$, the structure I was detected. 
The larger charges $Q_1$ constitute a slightly deformed hexagonal lattice,
see Fig. \ref{fig:StructureI}.
On the left side of this figure, each isosceles triangle has base of length
$a$ and two legs of length $\Delta_1 a$ with $\Delta_1\ge 1$.
The isosceles triangles on the right side have the same base of length
$a$ and two legs of length $\Delta_2 a$ with $\Delta_2\le 1$.
The energy minimization fixes the smaller charges $Q_2$ inside each of 
the left triangles, just on their symmetry axis (no vertical offset), at 
distance $d$ from the center of the triangle; 
altogether they form a zig-zag line.
In the limit $q\to 0$, it holds that $d\to 0$ and $\Delta_1, \Delta_2\to 1$, 
restoring the perfect hexagonal (equilateral triangle) lattice.

The periodic rectangle has the vertical side $A=a$ and the horizontal one 
$B = (\sqrt{\Delta_1^2-1/4}+\sqrt{\Delta_2^2-1/4}) a$.
The lattice constant $a$ can be expressed in terms of the particle density 
$\rho$ as follows.
There are two particles of each type per rectangle, with the total charge 
$2(Q_1+Q_2)$.
This charge compensates the surface charge density $\sigma$ within the
rectangle via $2 Q_1(1+q)=\sigma A B$, so using the relation (\ref{sigro})
we get for the reciprocal lattice constant
\begin{equation} \label{ai}
\frac{1}{a}=\frac{\sqrt{\rho}}{2}
\left( \sqrt{\Delta_1^2-1/4}+\sqrt{\Delta_2^2-1/4} \right)^{1/2} .
\end{equation}
The energy per particle of structure I is found to be
\begin{eqnarray} \label{e1}
&& \frac{E_{\rm I}(q;\Delta_1,\Delta_2,d)}{Q_1^2\sqrt{\rho}} = 
\frac{1}{4\sqrt{\pi}} \Bigg\{\frac{1}{2}\Sigma_1(\psi) 
\phantom{aaaaa} \nonumber \\ &&
+ \frac{1}{2} \Sigma_2\left(\psi,\frac{1}{\psi}\sqrt{\Delta_1^2-\frac{1}{4}},
\frac{1}{2}\right) \nonumber \\ &&
+q \Sigma_2\left[\psi,\frac{1}{\psi}\left(\frac{\Delta_1^2}{
\sqrt{4\Delta_1^2-1}}+d\right),0 \right] \nonumber \\ &&
+q \Sigma_2\left[\psi,\frac{1}{\psi}\left(\frac{\Delta_1^2-1/2}{
\sqrt{4\Delta_1^2-1}}-d\right),\frac{1}{2}\right]
+\frac{q^2}{2}\Sigma_1(\psi) \nonumber \\ && 
+\frac{q^2}{2}\Sigma_2\left[\psi,\frac{1}{\psi}\left(
-\frac{1}{2\sqrt{4\Delta_1^2-1}}+2d\right),\frac{1}{2}\right] \Bigg\},
\end{eqnarray}
where $\psi=B/A=\sqrt{\Delta_1^2-1/4}+\sqrt{\Delta_2^2-1/4}$.
Here, the terms of order $q^0$, $q^1$ and $q^2$ originate from 
$Q_1-Q_1$, $Q_1-Q_2$ and $Q_2-Q_2$ interactions, respectively.

\subsection{Structure II}

\begin{figure}[]
\begin{center}
\includegraphics[clip,width=0.34\textwidth]{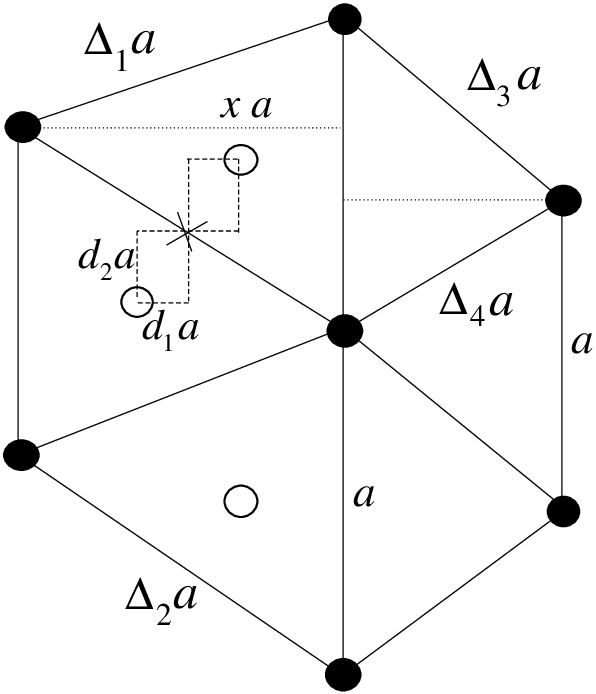}
\caption{Structure II. Full disks correspond to particles with charge $Q_1$, 
the empty ones to $Q_2$.  
The cross denotes the center of the longest side of left triangles.}
\label{fig:StructureII}
\end{center}
\end{figure}

Structure II resembles the structure I, but now the hexagon of particles 
with larger charge $Q_1$ is even more deformed, all sides of the triangles 
have different lengths: $\Delta_1 a$ and $\Delta_2 a$ on the left, 
$\Delta_3 a$ and $\Delta_4 a$ on the right, see Fig. \ref{fig:StructureII}.
The common vertical sides have length $a$.
The $Q_2$ charges are placed symmetrically with respect to the center of 
the longest sides $\Delta_2 a$.
Their coordinates measured from that center are $(d_1 a,d_2a)$ and 
$(-d_1 a,-d_2 a)$.
This choice of only two free parameters is supported by numerical simulations 
and mirror symmetry considerations.

Fig. \ref{fig:StructureII} does not show the whole periodic rectangle.
One must imagine another hexagon added, say on the right hand side, but with 
lengths of the sides reversed, i. e. $\Delta_1 \leftrightarrow \Delta_2$ and 
$\Delta_3 \leftrightarrow \Delta_4$.
The vertical side of the rectangle is still $A=a$, the horizontal one 
$B=\psi a$ is nontrivial, see below.
There are four particles of each type per rectangle and the reciprocal 
lattice constant is given by
\begin{eqnarray} \label{aii}
\frac{1}{a} & = & \frac{\sqrt{\rho}}{2\sqrt{2}}\left(
\sqrt{2\Delta_1^2+2\Delta_2^2+2\Delta_1^2\Delta_2^2-\Delta_1^4-\Delta_2^4-1}
\right. \nonumber \\ & & + \left. \sqrt{2\Delta_3^2+2\Delta_4^2+
2\Delta_3^2\Delta_4^2-\Delta_3^4-\Delta_4^4-1}\right)^{1/2} .
\end{eqnarray}
Note that this structure can continuously go over into the structure I, 
if $\Delta_1=\Delta_2$, $\Delta_3=\Delta_4$ and $d_2=1/4$.
The energy of structure II reads
\begin{widetext}
\begin{eqnarray} \label{e2}
&&\frac{E_{\rm II}(q;\Delta_1,\Delta_2,\Delta_3,\Delta_4,d_1,d_2)}{
Q_1^2\sqrt{\rho}} = \frac{1}{4\sqrt{2\pi}} \Bigg\{
\frac{1+q^2}{2}\Sigma_1(\psi) + \frac{1}{2} \Bigg[
\Sigma_2\left(\psi,\frac{x}{\psi},\frac{\Delta_1^2-\Delta_2^2+1}{2}\right) + 
\Sigma_2\left(\psi,\frac{1}{2}+\frac{x}{\psi},\frac{\Delta_4^2-\Delta_3^2+1}{2}
\right) \nonumber \\ &&
+ \Sigma_2\left(\psi,\frac{1}{2},\frac{\Delta_1^2-\Delta_2^2-\Delta_3^2
+\Delta_4^2}{2}\right)\Bigg] +
q\bigg[\Sigma_2\left(\psi,\frac{x}{2\psi}+d_1,
\frac{\Delta_1^2-\Delta_2^2-1}{4}+d_2 \right) + 
\Sigma_2\left(\psi,\frac{x}{2\psi}-d_1,\frac{\Delta_1^2-\Delta_2^2-1}{4}-d_2 
\right) \nonumber \\ && +
\Sigma_2\left(\psi,\frac{1}{2}+\frac{x}{2\psi}-d_1,
\frac{\Delta_1^2-\Delta_2^2+1}{4}-\frac{\Delta_3^2-\Delta_4^2}{2}+d_2 \right) +
\Sigma_2\left(\psi,\frac{1}{2}+\frac{x}{2\psi}+d_1,
\frac{\Delta_1^2-\Delta_2^2+1}{4}-\frac{\Delta_3^2-\Delta_4^2}{2}-d_2 
\right) \Bigg] \nonumber \\ && +
\frac{q^2}{2} \left[ \Sigma_2\left(\psi,-2 d_1,-2 d_2 \right) +
\Sigma_2\left( \psi,\frac{1}{2}-2 d_1,\frac{\Delta_4^2-\Delta_3^2+1}{2} \right)
+ \Sigma_2\left(\psi,\frac{1}{2},\frac{\Delta_4^2-\Delta_3^2+1}{2}-2 d_2 \right)
\right] \Bigg\} .
\end{eqnarray}
The altitude $x a$ of the left triangle in Fig. \ref{fig:StructureII} 
is given by
\begin{equation}
x = \frac{1}{2} 
\sqrt{2\Delta_1^2+2\Delta_2^2+2\Delta_1^2\Delta_2^2-\Delta_1^4-\Delta_2^4-1}.
\end{equation}
A similar expression holds for the right triangle.
Consequently, $\psi=B/A$ reads as
\begin{equation}
\psi = 
\sqrt{2\Delta_1^2+2\Delta_2^2+2\Delta_1^2\Delta_2^2-\Delta_1^4-\Delta_2^4-1} +
\sqrt{2\Delta_3^2+2\Delta_4^2+2\Delta_3^2\Delta_4^2-\Delta_3^4-\Delta_4^4-1} .
\end{equation}
\end{widetext}

\subsection{Structures III - V}
The most simple complete mixture is structure III.
It is composed of two intertwining square lattices with the same lattice 
constant $a$.
The larger $Q_1$ charges occupy the vertices of one square lattice, 
the $Q_2$ charges occupy the other one shifted by vector $(a/2,a/2)$. 
The parameter $\psi = B/A = 1$.
There is one particle of each type per square.
The reciprocal lattice constant is given by $1/a=\sqrt{\rho/2}$.
For the energy per particle, we get the simple formula
\begin{equation} \label{e3}
\frac{E_{\rm III}(q)}{Q_1^2\sqrt{\rho}} = \frac{1}{2\sqrt{2\pi}}
\left[ \frac{1+q^2}{2}\Sigma_1(1)
+ q \Sigma_2\left(1,\frac{1}{2},\frac{1}{2}\right) \right].
\end{equation}
This phase has no internal parameters, so no optimization is needed.

We mention structures IV and V only briefly.\cite{Ant}
They appear in a narrow interval of $q$ values, namely 
$0.26\lesssim q \lesssim 0.28$. 
They are characterized by a large number of internal parameters and one 
cannot go smoothly neither between themselves, nor to the neighboring III 
and VI structures, hence all transitions must be of first order.
Due to the absence of interesting phenomena, we decided not to calculate 
structures IV and V by our method and in the phase diagram we shall use 
values got numerically by Antlanger and Kahl.\cite{Ant}

\subsection{Structure VI}

\begin{figure}[]
\begin{center}
\includegraphics[clip,width=0.4\textwidth]{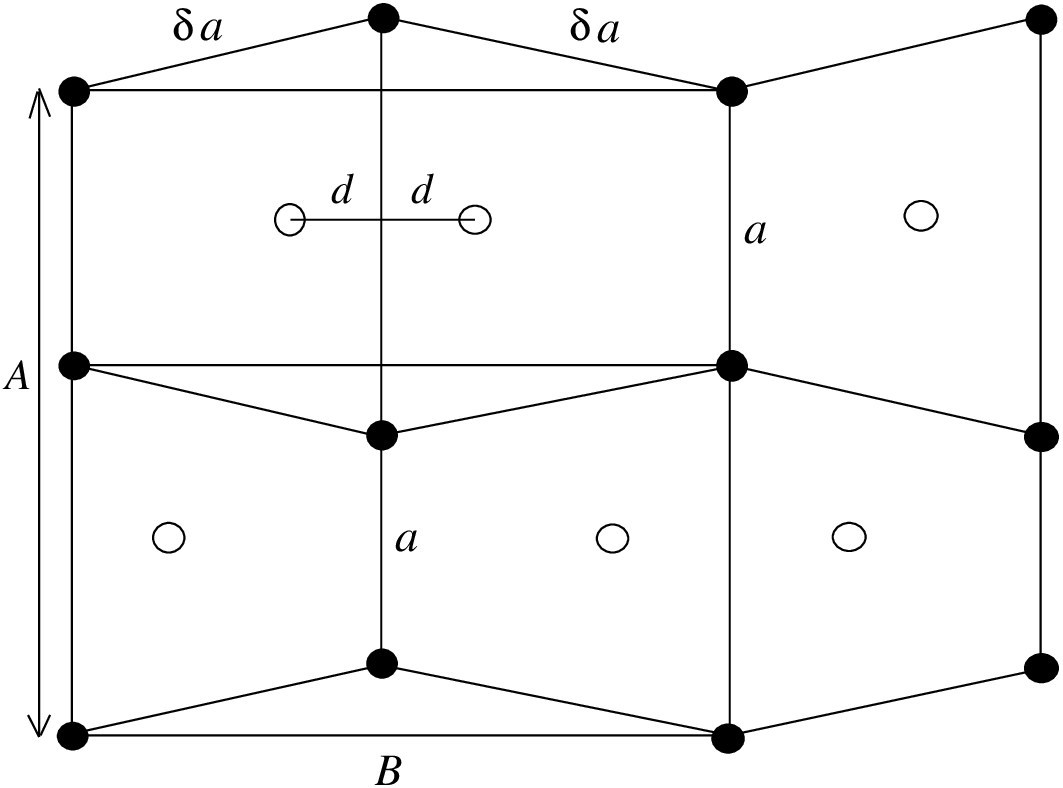}
\caption{Structure VI. Full disks correspond to particles with charge $Q_1$, 
the empty ones to $Q_2$.}  
\label{fig:StructureVI}
\end{center}
\end{figure}

Structure VI is presented in Fig. \ref{fig:StructureVI}. 
The charges $Q_1$ occupy hexagons of a deformed honeycomb lattice with
two sides of length $a$ and four sides of length $\delta a$; $\delta$
varies over a small interval $[0.959,1.0054]$. 
Two smaller charges $Q_2$ enter the interior of each hexagon and are placed 
symmetrically at horizontal distances $\pm d$ with respect to 
the hexagon center. 
The sides $A$ and $B$ of the basic periodic rectangle are marked in the figure. 
The horizontal side $B=\Delta a$, where $\Delta=\sqrt{3}$ for a fully 
symmetric hexagon and we have theoretical bounds $\sqrt{3}<\Delta<2$ for 
a deformed hexagon.
The vertical side $A$ is easily deduced in the form
\begin{equation}
A = 2 \left( 1+\sqrt{\delta^2-\frac{\Delta^2}{4}} \right) a .
\end{equation}
There are four particles of each type per rectangle.
The reciprocal lattice constant is found to be
\begin{equation} \label{avi}
\frac{1}{a} = \frac{\sqrt{\rho}}{2}\left[ \Delta
\left( 1+\sqrt{\delta^2-\frac{\Delta^2}{4}} \right) \right]^{1/2} .
\end{equation}
The energy is obtained in the form
\begin{widetext}
\begin{eqnarray} \label{e6}
\frac{E_{\rm VI}(q;\Delta,\delta,d)}{Q_1^2\sqrt{\rho}} & = &
\frac{1}{4\sqrt{2\pi}} \Biggl( \frac{1+q^2}{2}\Sigma_1(\psi)+
\frac{1}{2} \Sigma_2\left( \psi,0,\frac{1}{A'}\right) +\frac{1}{2}
\Sigma_2\left( \psi,\frac{1}{2},-\frac{\sqrt{\delta^2-\Delta^2/4}}{A'}\right)
+\frac{1}{2}\Sigma_2\left(\psi,\frac{1}{2},\frac{1}{2}\right) \nonumber \\
& & + q \Biggl\{ \Sigma_2\left(\psi,\frac{1}{2}-d,\frac{1}{2A'}\right)
+\Sigma_2\left( \psi,\frac{1}{2}+d,\frac{1}{2A'}\right) \nonumber \\
& & + \Sigma_2\left[ \psi,-d,-\frac{1}{A'}
\left( \frac{1}{2}+\sqrt{\delta^2-\frac{\Delta^2}{4}}\right) \right]
+ \Sigma_2\left[ \psi,d,-\frac{1}{A'}
\left( \frac{1}{2}+\sqrt{\delta^2-\frac{\Delta^2}{4}}\right) \right] \Biggr\}
\nonumber \\ & & + \frac{q^2}{2} \left[ \Sigma_2(\psi,2d,0)
+ \Sigma_2\left( \psi,\frac{1}{2},\frac{1}{2}\right)
+ \Sigma_2\left( \psi,\frac{1}{2}+2d,\frac{1}{2}\right) \right] \Biggr) ,
\end{eqnarray}
where $A'\equiv A/a = 2(1+\sqrt{\delta^2-\Delta^2/4})$ and $\psi=B/A=\Delta/A'$.
\end{widetext}

\subsection{Separation of pure phases}
For large enough $q\gtrsim 0.59$, the energy of the nontrivial structure
VI becomes higher than that of the pure phase separation in which the charged 
species $Q_1$ and $Q_2$ occupy spatially separated infinite areas, each in 
a hexagonal arrangement (which is known to give the energy minimum for 
the planar Coulomb particles of one type) immersed in a neutralizing 
background $\sigma$, with the lattice constant given by the requirement of 
electroneutrality. 

First we sketch the formula for the energy per particle of charges $Q$, 
localized on vertices of the hexagonal lattice 
and immersed in a neutralizing background $\sigma$. 
There is one particle per two triangles, each with area 
$S_{\Delta}=a^2\sqrt{3}/4$ where $a$ is the lattice constant.
The electroneutrality requires that $Q = 2\sigma S_\Delta$, so that 
\begin{equation} \label{aqs}
a = \sqrt{\frac{2Q}{\sqrt{3}\sigma}} .
\end{equation}
The distance of two interacting particles is proportional to $a$ and the 
energy per particle including the neutralizing background energy is\cite{Samaj}
\begin{equation}\label{ehex}
E_{\rm hex}(Q) = Q^2\sum_{j\ne 1} \frac{1}{r_{1j}} + {\rm backgr}
= c_M \sqrt{\sigma} Q^{3/2} ,
\end{equation}
where $c_{\rm M} =-1.96051578931989165120575262921\ldots$ is 
the Madelung constant of the hexagonal lattice. 

For the separation of the pure phases of $N/2$ particles with charge $Q_1$ 
and $N/2$ particles with charge $Q_2$, each constituting the corresponding 
hexagonal structure, the total energy $E_0^{\rm T}(q)$ is\cite{Ant} 
\begin{eqnarray} 
E_0^{\rm T}(q) & = & \frac{N}{2} \left[ E_{\rm hex}(Q_1)+E_{\rm hex}(Q_2) \right]
\nonumber \\ & = &
\frac{N}{2} \sqrt{\sigma} c_{\rm M} \left( Q_1^{3/2}+Q_2^{3/2} \right) . \label{et0}
\end{eqnarray}
With regard to Eq. (\ref{sigro}), the energy per particle 
$E_0(q) = E_0^{\rm T}(q)/N$ is given by
\begin{equation} \label{e0}
\frac{E_0(q)}{Q_1^2\sqrt{\rho}} = c_{\rm M} 
\sqrt{\frac{1 + q}{2}} \frac{1 + q^{\frac{3}{2}}}{2} .
\end{equation}

As was mentioned above, the separation is energetically favorable for higher 
values of $q$, but its energy will serve as a reference for any $q$.

\section{Infinite hierarchy of separated pure and mixed phases} \label{Sec3}
\begin{figure}[] 
\begin{center}
\includegraphics[clip,width=0.4\textwidth]{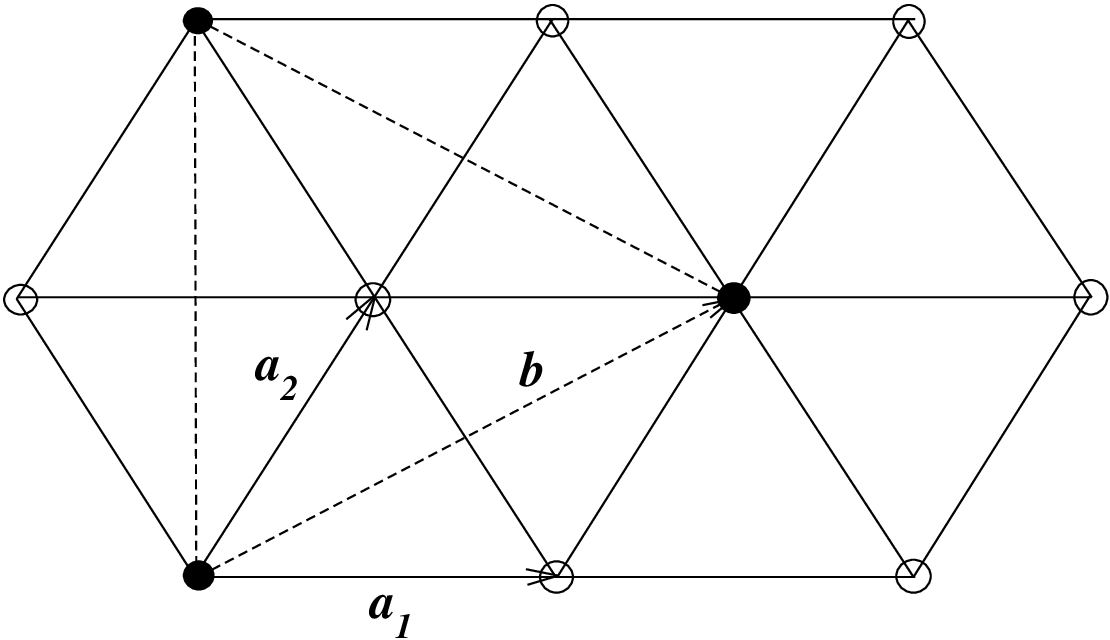}
\caption{The mixed phase $B$ of the separation \{1,1\}. 
Full disks correspond to particles with charge $Q_1$, the empty ones to $Q_2$.  
The basic hexagonal structure (solid lines) has lattice constant $a$, 
particles $Q_1$ form another hexagonal lattice with lattice constant 
$b = \sqrt{3} a$.}
\label{fig:1-1}
\end{center}
\end{figure}

Now we introduce a novel structure which consists in the separation
of a pure $A$ and a mixed $B$ phases.
The pure phase $A$ is composed of $N_1^A< N/2$ charges $Q_1$ in a hexagonal 
arrangement; there are no $Q_2$ charges in phase $A$, i.e. $N_2^A=0$.
The mixed phase $B$ is composed of the remaining $N_1^B = N/2-N_1^A$ charges 
$Q_1$ and all $N_2^B=N/2$ charges $Q_2$. 
The charges are distributed in a specific way within a basic hexagonal lattice. 
We denote the primitive translation vectors of this basic hexagonal lattice as 
\begin{equation}
{\bm a}_1 = a(1,0) , \qquad {\bm a}_2 = \frac{a}{2} (1,\sqrt{3}) ,
\end{equation}
see Fig. \ref{fig:1-1}.
Let the vector ${\bm b}$ be defined as
\begin{equation} \label{vecb}
{\bm b} = j {\bm a}_1 + k {\bm a}_2, \quad
b^2 = a^2 (j^2+jk+k^2),
\end{equation}
where $\{j,k\}$ are non-negative integers, at least one of them 
being nonzero, i.e. $\{0,2\}$, $\{1,1\}$, $\{0,3\}$, $\{1,2\}$, etc.
For a given $\{j,k\}$, charges $Q_1$ occupy a hexagonal subset of the basic 
hexagonal lattice with the lattice constant $b$; it is a special property
of any hexagonal lattice $\alpha$ that joining arbitrary two vertices of 
$\alpha$ implies a side of the new hexagonal lattice $\beta$ whose all 
points also belong to $\alpha$.    
The remaining sites are occupied by charges $Q_2$.
The special case $\{1,1\}$ is pictured in Fig. \ref{fig:1-1}.
The particle densities of fully occupied hexagonal structures are proportional 
to the reciprocal of the squared lattice constant.
In particular, $\rho_1^B\propto 1/b^2$ and $\rho_{1+2}^B\propto 1/a^2$ since 
the the basic hexagonal lattice contains both types of particles.
Consequently,
\begin{equation} \label{n1b}
\frac{\rho_1^B}{\rho_{1+2}^B} = \frac{N_1^B}{N_2^B+N_1^B}
= \frac{a^2}{b^2} = \frac{1}{j^2+jk+k^2} . 
\end{equation}
Using the fact that $N_2^B = N/2$, the number of $Q_1$ particles in phase $B$ 
is equal to
\begin{equation}
N_1^B = \frac{N}{2} \frac{1}{j^2+jk+k^2-1} .
\end{equation}

To calculate the total energy for the pure hexagonal phase $A$ with 
\begin{equation} 
N_1^A = \frac{N}{2}-N_1^B=\frac{N}{2}\frac{j^2+jk+k^2-2}{j^2+jk+k^2-1}
\end{equation}
particles of charge $Q_1$, we use Eq. (\ref{ehex}) and obtain the result
\begin{equation} \label{rega}
E_A^{\rm T} = c_{\rm M} \sqrt{\sigma} Q_1^{3/2}
\frac{N}{2} \frac{j^2+jk+k^2-2}{j^2+jk+k^2-1} .
\end{equation}

To calculate the total energy for the mixed phase $B$, we introduce the average 
charge $\bar{Q}$ as follows
\begin{equation} \label{barq}
\bar{Q} = \frac{N_1^BQ_1+N_2^BQ_2}{N_1^B+N_2^B}
= Q_1\frac{1+(j^2+jk+k^2-1)q}{j^2+jk+k^2} .
\end{equation}
In analogy with Eq. (\ref{aqs}), the lattice constant $a$ is given by
\begin{equation} \label{aqsbar}
a = \sqrt{\frac{2\bar{Q}}{\sqrt{3}\sigma}} .
\end{equation}
We apply a trick which consists in splitting each charge $Q_1$ into 
two charges, $Q_1=Q_2+(Q_1-Q_2)$, and then performing three summations 
over hexagonal lattices only.
The first contribution sums the mutual interactions of $N/2+N_1^B$ charges 
$Q_2$ on the hexagonal lattice with lattice constant $a$,
\begin{equation}
c_{\rm M} \sqrt{\sigma} \frac{Q_2^2}{\sqrt{\bar{Q}}} 
\left( \frac{N}{2} + N_1^B \right) .
\end{equation}
The second one sums the interactions of $N_1^B$ $(Q_1-Q_2)$ charges with all 
other charges $Q_2$ on the hexagonal lattice with lattice constant $a$,
\begin{equation}
2 c_{\rm M} \sqrt{\sigma} \frac{Q_2 (Q_1-Q_2)}{\sqrt{\bar{Q}}} N_1^B .
\end{equation}
Note the factor 2 for the interaction of different species.
The third contribution sums mutual interactions of $N_1^B$ charges $(Q_1-Q_2)$
on the hexagonal lattice with lattice constant $b$, 
\begin{equation}
c_{\rm M} \sqrt{\sigma} \frac{(Q_1-Q_2)^2}{\sqrt{\bar{Q}}} \frac{a}{b} N_1^B .
\end{equation}
To get the total energy of the phase separation, coined by $\{ j,k\}$,
we sum all three terms with the energy (\ref{rega}) of the pure phase $A$. 
The final formula for the energy per particle becomes
\begin{eqnarray} \label{ejk}
\frac{E(q; j, k)}{Q_1^2 \sqrt{\rho}} = \frac{c_{\rm M}}{2} \sqrt{\frac{1 + q}{2}}
\Bigg[\frac{j^2 + j\ k + k^2 - 2}{j^2 + j\ k + k^2 - 1} \phantom{aaaaa}
\nonumber\\ + q^2\frac{j^2 + j\ k + k^2}{j^2 + j\ k + k^2 - 1}
\sqrt{\frac{j^2 + j\ k + k^2}{1 + (j^2 + j\ k + k^2 - 1)q}} \nonumber\\
     + \frac{2q(1 - q)}{j^2 + j\ k + k^2 - 1}
     \sqrt{\frac{j^2 + j\ k + k^2}{1 + (j^2 + j\ k + k^2 - 1)q}} \nonumber\\
+ \frac{(1 - q)^2}{j^2 + j\ k + k^2 - 1}
   \sqrt{\frac{1}{1 + (j^2 + j\ k + k^2 - 1)q}} \Bigg] .
\end{eqnarray}
Note that the limit $j\to\infty$ or $k\to\infty$ of this relation
reproduces correctly Eq. (\ref{e0}).

The separation $\{1,1\}$ with the mixed $B$ phase presented in 
Fig. \ref{fig:1-1} is special: each charge $Q_1$ is in the middle of 
an hexagon formed by six $Q_2$ charges and each charge $Q_2$ is in 
the middle of a triangle formed by three $Q_1$ charges.
Due to this symmetry, the total force acting on particles vanishes
and the particles are in the true ground state with the energy per particle
given by Eq. (\ref{ejk}) as follows
\begin{eqnarray} \label{e11}
\frac{E(q;1,1)}{Q_1^2\sqrt{\rho}} = \frac{c_{\rm M}}{2} \sqrt{\frac{1 + q}{2}}
\Bigg[ \frac{1}{2} + \frac{3}{2} q^2 \sqrt{\frac{3}{1+2q}} 
\phantom{aaa} \nonumber \\
+ q (1-q) \sqrt{\frac{3}{1+2q}} + \frac{(1-q)^2}{2} \frac{1}{\sqrt{1+2q}}
\Bigg] . 
\end{eqnarray}

\begin{figure}[] 
\begin{center}
\includegraphics[clip,width=0.37\textwidth]{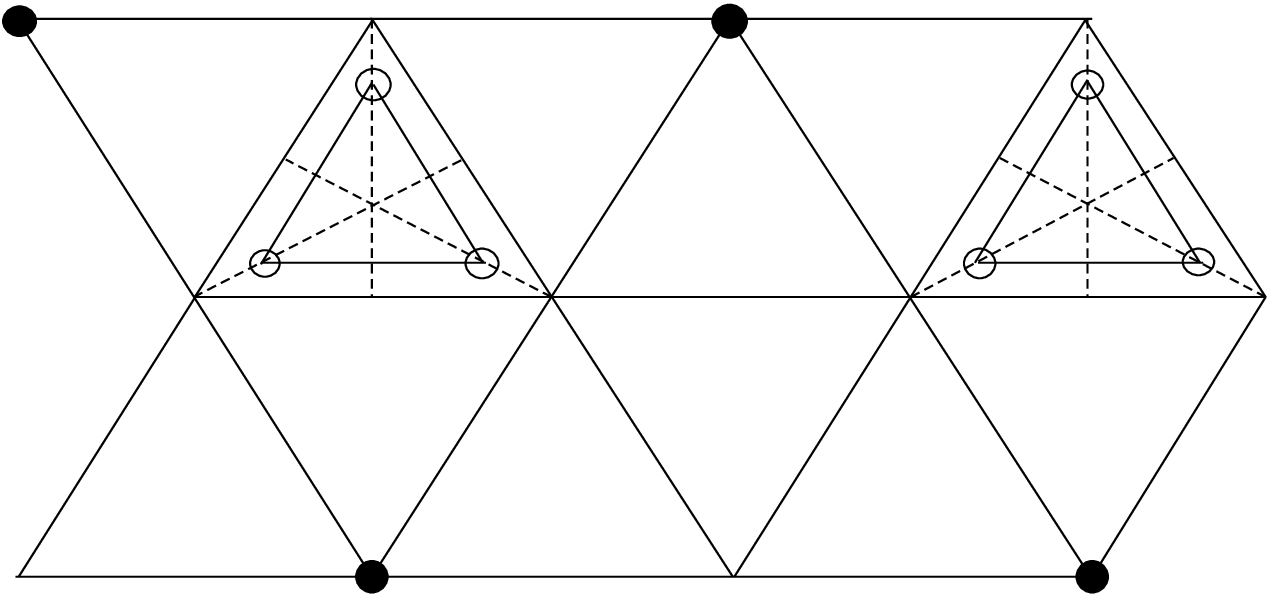}
\caption{Relaxed mixture phase $B$ of the separation $\{0,2\}$. 
Full disks correspond to particles with charge $Q_1$, the empty ones to $Q_2$.
There are three $Q_2$ particles within each rhombus formed by $Q_1$ vertices.}
\label{fig:0-2}
\end{center}
\end{figure}

For other separations $\{j,k\}$, the particles are in general in unstable
positions and should relax to the optimal ground-state positions.
For the structure $\{0,2\}$ in Fig. \ref{fig:0-2}, let us make the lattice 
of (black) $Q_1$ charges rigid.
The (white) $Q_2$ charges experience an asymmetric force with a saddle
point at the center of the rhombus.
Fig. \ref{fig:0-2} documents one of possibilities how $Q_2$ charges 
can relax within the equilateral triangles. 
For small $q$, the distance of $Q_2$ charges to $Q_1$ increases, since 
its reciprocal enters the energy term linear in $q$, whereas the mutual 
$Q_2-Q_2$interactions are only of the order $q^2$, allowing them to get closer.
Symmetry considerations suggest that the distance to each pair of $Q_1$ 
particles should be equal, which is represented by the dashed lines and 
the $Q_2$ particles form a smaller equilateral triangle.
For these relaxed configurations, we indeed get slightly lower energies
than for the non-relaxed ones.
There are other possibilities, but we do not go into details since for any
$q$ all these relaxed structure have higher energies than the simplest rigid 
$\{1,1\}$ structure.

\section{Results} \label{Sec4}
We have used the above formulas for various structures to optimize 
the minimum energy in the whole interval $0\le q \le 1$. 
The approach taken from our former works\cite{Samaj,TS} allowed us to achieve 
at least 15 decimal digits precision, compared to typical 7 decimal digits 
in numerics.\cite{Ant}
We included the novel separated $\{ 1,1\}$ phase which turns out to have
the lowest energy both in the region of small and intermediate values of $q$.
The differences in energies between competing structures are often very small, 
hence we will subtract the reference value $E_0(q)$, see Eq. (\ref{e0}), 
which makes them more visible. 
The quantity in our figures will be $[E(q)-E_0(q)]/(\sqrt{\rho}Q_1^2)$, 
where $E(q)$ is the energy of the considered structure.
It stands to reason that when the separation of the two pure phases becomes 
dominant, the corresponding energy difference is 0. 

\subsection{Small values of $q$}
\begin{figure}[]
\begin{center}
\includegraphics[clip,width=0.45\textwidth]{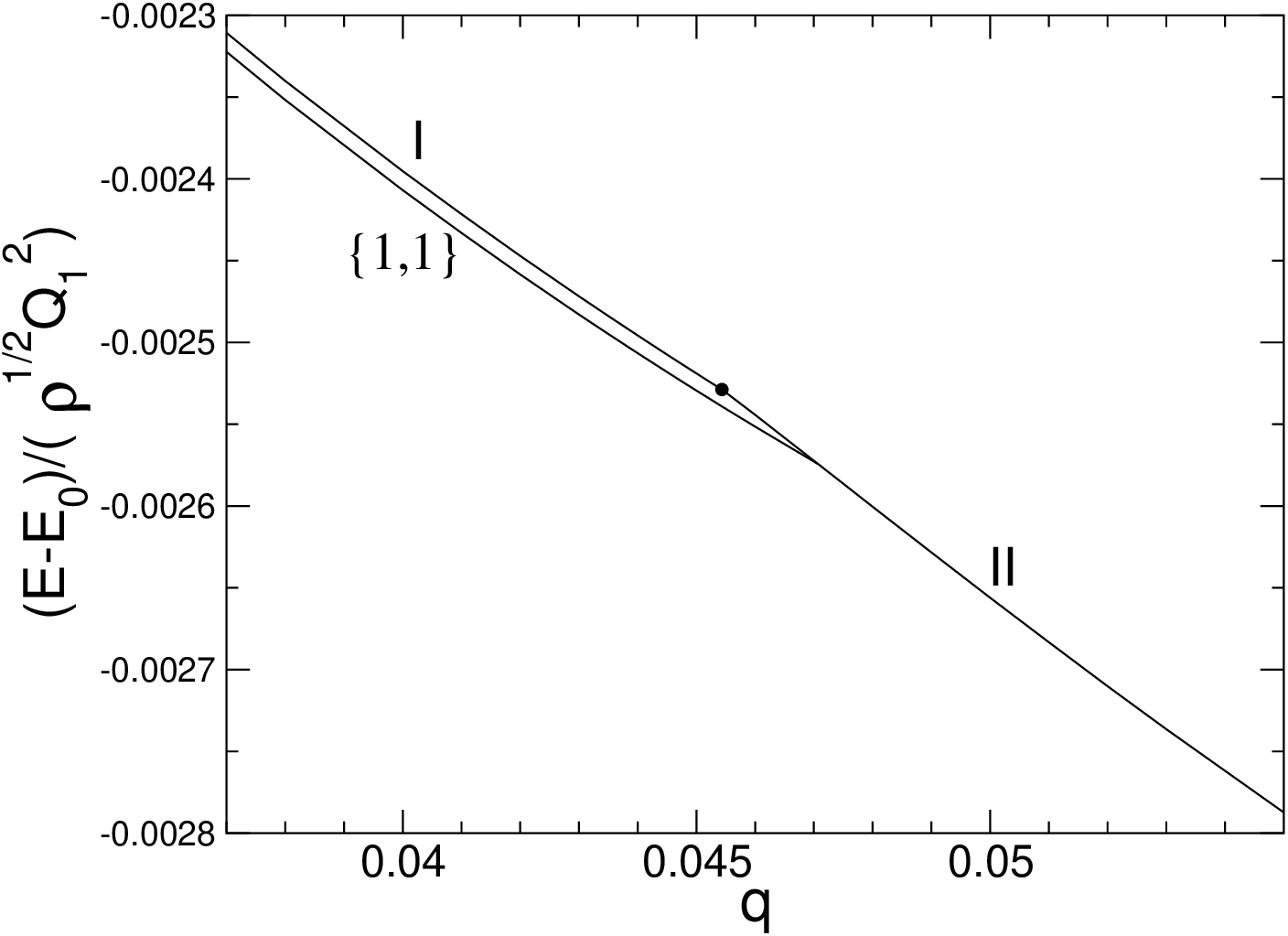}
\caption{The energy difference $[E(q)-E_0(q)]/(\sqrt{\rho}Q_1^2)$ 
for completely mixed structures I, II and for the separation of the pure and 
mixed phases $\{ 1,1\}$ in the region of small values of $q$. 
The transition point between structures I and II is depicted by the full disc.
This point is ineffective since the structure $\{ 1,1 \}$ is dominant 
well beyond it.
Structure II becomes the one with the lowest energy at $q\approx 0.04707$.}
\label{fig:Small-q}
\end{center}
\end{figure}

It was claimed\cite{Ant} that the energy $E_{\rm I}$ of structure I gives 
the lowest energy for small values of $q$, up to $q\approx 0.046$ when 
structure II takes the place. 
We found that the energy of the separated (pure and mixed) structure 
$E(q;1,1)$ is lower than $E_{\rm I}$ in the whole interval of structure I 
dominance. 
It is even lower than $E_{\rm II}$ within a short interval of $q$-values, 
see Fig. \ref{fig:Small-q}.

The difference between $E_{\rm I}(q)$ and $E(q;1,1)$ is tiny, of 
order $10^{-5}$ which is on the border of accuracy of numerical methods.
In this work, we take advantage of analytic formulas and calculate 
the small-$q$ expansions of both the optimized $E_{\rm I}(q)$ and $E(q;1,1)$.
From the representation (\ref{e1}), we deduce the expansion
\begin{eqnarray} \label{e1exp}
E_{\rm I}(q;\Delta_1,\Delta_2,d) \approx E_0(0) 
\Big[ 1+\left( \sqrt{3}-1 \right) q \phantom{aaaa} \nonumber\\
+ 1.3428568510125 q^2 - 0.300557 q^3 \Big] + {\cal O}\left( q^4 \right) 
\end{eqnarray}
with $E_0(0)/(Q_1^2\sqrt{\rho}) = c_{\rm M}/(2\sqrt{2})\approx -0.6931470046$ 
being negative.
Here, the coefficients up to the $q^2$ term were derived analytically in 
Appendix \ref{ApC}, the coefficient at $q^3$ comes from numerical fittings.
This expression should be compared to the analogous expansion of 
Eq. (\ref{e11}):
\begin{eqnarray} \label{e11exp}
E(q;1,1) & = & E_0(0) \left[ 1+(\sqrt{3}-1)q +\frac{11}{8}q^2 \right. 
\nonumber \\ & & \left. - \frac{5}{8} (3-\sqrt{3}) q^3 \right] 
+{\cal O}\left( q^4\right) \nonumber\\ & = & E_0(0) 
\Big[ 1+(\sqrt{3}-1)q +1.375q^2 \nonumber \\ & & 
- 0.79246825 q^3 \Big] +{\cal O}\left( q^4\right) .
\end{eqnarray}
The two expansions (\ref{e1exp}) and (\ref{e11exp}) coincide up to the term
linear in $q$.
Since $E_0(0)<0$ and the prefactor to the $q^2$ term in (\ref{e11exp}) is 
larger than that in (\ref{e1exp}), the separation of the pure and mixed phases
$\{ 1,1\}$ is dominant in the region of small $q$.
Numerically one finds that $E(q;1,1)<E_{\rm I}(q)$ for $q\le 0.05$. 

We mentioned the possibility that the structure I goes smoothly into 
structure II via a second-order phase transition.
If our separation $\{1,1\}$ is not taken into account, the transition between
structures I and II at $q=0.045434$ would be of first order, i.e. there is 
a change in the slope of the upper line in Fig. \ref{fig:Small-q} and 
the structural parameters change discontinuously.
But the structure I is wiped out by the separation $\{1,1\}$ and 
the latter performs a first-order transition directly into the structure II 
at $q=0.04707$.

The next transition between structures II and III takes place at $q=0.09125$.

\subsection{Intermediate values of $q$}

\begin{figure}[]
\begin{center}
\includegraphics[clip,width=0.45\textwidth]{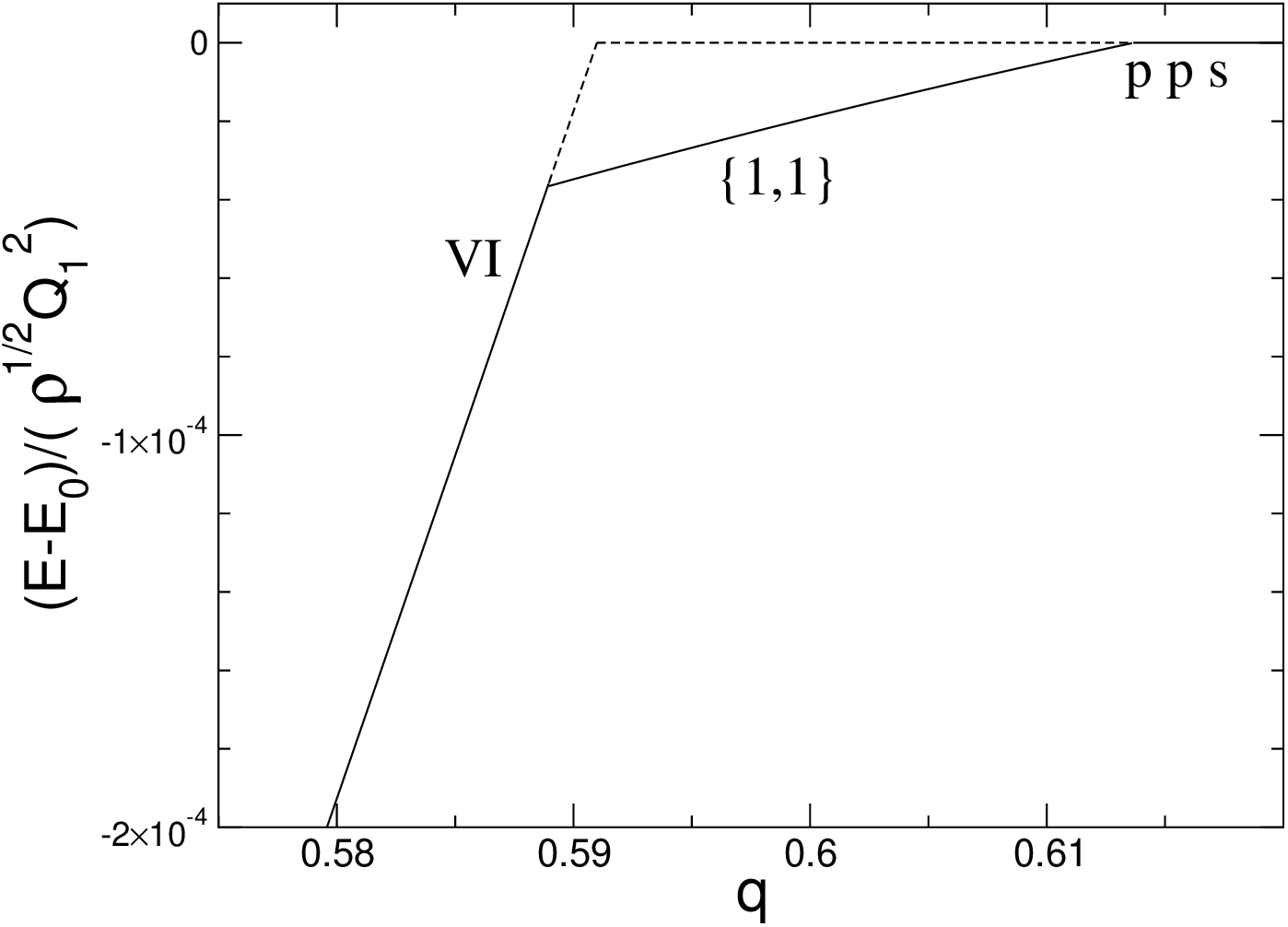}
\caption{$[E(q)-E_0(q)]/(\sqrt{\rho}Q_1^2)$ for intermediate values of $q$. 
The interplay among mixed phase VI, our separation of pure and mixed phases
$\{ 1,1\}$ and the separation of pure phases $E_0$.
The dashed lines would be valid if the separation $\{1,1\}$ is ignored.
The abbreviation ``pps'' means ``pure phase separation'', 
its energy $E_0(q)$ serving as a reference.}
\label{fig:Medium-q}
\end{center}
\end{figure}

It was found\cite{Ant} that the completely mixed structure VI goes over 
into the separation of two pure phases, represented by the zero value 
of $E_0-E_0$, via a first-order transition at $q=0.59099$.
We found surprisingly that the separation $\{1,1\}$ provides the lowest 
energy also for intermediate values of $q$, namely for $0.58895<q<0.61367$,
see Fig. \ref{fig:Medium-q}.
The lower value of $q=0.58895$ marks the first-order transition between 
structures VI and $\{1,1\}$, the higher $q=0.61367$ the first-order
transition between $\{ 1,1\}$ and the separation of two pure phases.
Note extremely small differences among energies.

\subsection{Phase diagram for all values of $q$}
\begin{figure}[]
\begin{center}
\includegraphics[clip,width=0.45\textwidth]{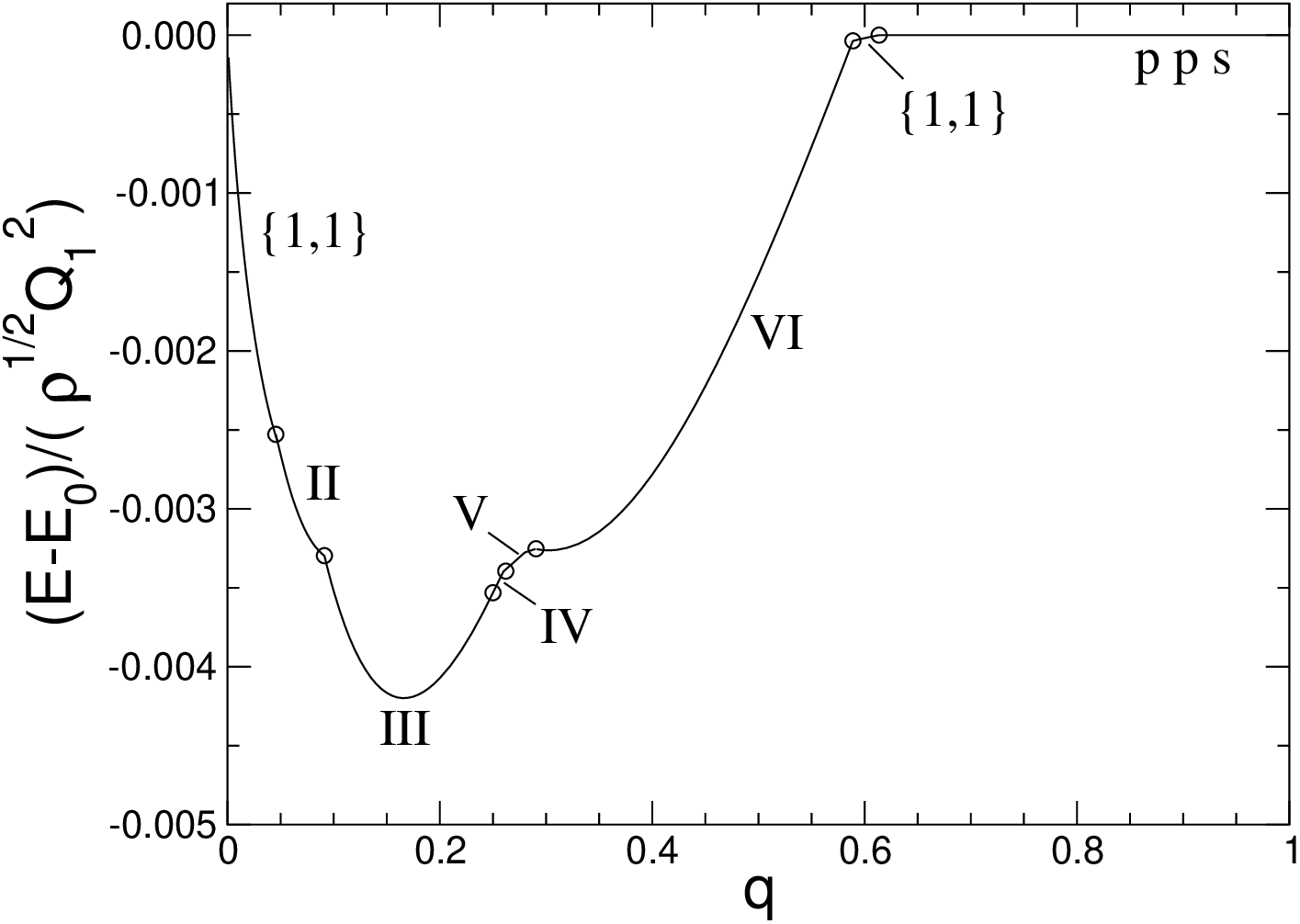}
\caption{$[E(q)-E_0(q)]/(\sqrt{\rho}Q_1^2)$ for all values of $q$. 
Open circles mark the first-order transition points.
The abbreviation ``pps'' means ``pure phase separation''.}
\label{fig:All-q}
\end{center}
\end{figure}

Finally, we summarize the minimum energies of dominant phases for all values 
of $0\le q\le 1$ in Fig. \ref{fig:All-q}.
The energies are calculated by using the series formulas given above, with 
the exception of marginal phases IV and V; the corresponding data 
were taken from the previous numerical study.\cite{Ant} 
The open circles denote the first-order transition points.
Note that the separated structures emerge when $Q_2$ particles are almost 
negligible or comparable to $Q_1$.
The phase separation of pure hexagonal phases was observed also 
in the case of Yukawa interaction,\cite{AML} for $q\gtrsim 0.59$. 

\section{Conclusions} \label{Sec5}
We have shown that besides the perfectly mixed structures and the separated 
pure phase, a special separation of pure and mixed phases $\{ 1,1\}$ 
can appear in the phase diagram of a planar, equimolar binary mixture 
of classical charged particles. 
The energy of the separation structure $\{1,1\}$ turns out to be the lowest 
one in two regions of $q$: it dominates the completely mixed I and II phases 
for $0<q\lesssim 0.04707$ and the completely mixed VI phase and the separation
of two pure phase for $0.58895\lesssim q\lesssim 0.61367$.
The energy differences among the considered phases are very small. 
We obtained the results by using an analytic method based on an expansion of 
the Coulomb interaction energy in Misra functions\cite{Samaj,TS} which
enables us to get the energies with an accuracy by a dozen digits better than 
that of EA numerical simulations.\cite{Ant}

As concerns possible variations of the model, one could consider different
numbers of two particle species, like it was done in Monte Carlo approach
of Yukawa binary mixtures.\cite{AML} 
Another interesting problem is to confine the binary Coulomb mixture
between two (symmetrically or asymmetrically) charged plates; with regard
to the planar case, there should exist a larger variety of bilayer Wigner 
crystals.
Our method permits one to study inverse-power-law or Lenard-Jones 
particles\cite{Schweigert} as well.

This paper might inspire physicists oriented in numerical computation to test 
more general separations of binary mixtures, going beyond a simple separation 
of two pure phases.
We have proposed a specific separation of a pure and a mixed phase which 
provide the lowest energy in comparison with the known structures 
in two intervals in the regions of small and intermediate values of
parameter $q$.
But there may exist other separations, say of two or more completely mixed 
phases, with even lower energy.
It is not clear how to test numerically all possibilities on a reasonable 
time scale.

\begin{acknowledgments}
We would like to thank Moritz Antlanger and Gerhard Kahl for 
sending us their numerical data of energy,\cite{Ant} partially used in 
Fig. \ref{fig:All-q}.
The support received from the grant VEGA No. 2/0015/2015 is acknowledged. 
\end{acknowledgments}

\appendix

\section{Misra functions} \label{ApA}
The Misra functions are defined by\cite{Misra}
\begin{equation} \label{Misra}
z_{\nu}(y) = \int_0^{1/\pi} \frac{dt}{t^{\nu}} \exp\left( -\frac{y}{t} \right) ,
\quad y>0 .
\end{equation}
In this paper, we use the Misra functions with the half-integer argument
$\nu$ which are expressible in terms of the complementary error 
function\cite{Gradshteyn}
\begin{equation} \label{erfc}
{\rm erfc}(z)=\frac{2}{\sqrt{\pi}}\int_z^\infty\exp{(-t^2)}\ {\rm d}t.
\end{equation}
For the first Misra functions, we have\cite{TS}
\begin{eqnarray}
z_{3/2}(y) & = & \sqrt{\frac{\pi}{y}}\ {\rm erfc}{\left(\sqrt{\pi y}\right)},
\nonumber \\
z_{5/2}(y) & = & \frac{\sqrt{\pi}}{2 y^{3/2}}\left[ 2 {\rm e}^{-\pi y}\sqrt{y} 
+{\rm erfc}{\left(\sqrt{\pi y}\right)} \right] , \nonumber \\
z_{7/2}(y) & = & \frac{\sqrt{\pi}}{4 y^{5/2}}
\left[ 2 {\rm e}^{-\pi y}\sqrt{y}\left(3+2\pi y\right) 
+ 3\ {\rm erfc}{\left(\sqrt{\pi y}\right)} \right] . 
\nonumber \\ & & \label{znu0y}
\end{eqnarray}

For a small $\delta y$, we shall need the following expansion
\begin{eqnarray}
z_{\nu}\left(y+\delta y\right) & = & 
\int_0^{1/\pi} \frac{{\rm d}t}{t^\nu}{\rm e}^{-(y+\delta y)/t} \nonumber\\
&\approx & \int_0^{1/\pi} \frac{{\rm d}t}{t^\nu}{\rm e}^{-y/t}
\left[ 1-\delta y/t+\frac{(\delta y)^2}{2 t^2} \right] \nonumber\\
& = & z_\nu(y) -\delta y\ z_{\nu+1}(y) + \frac{(\delta y)^2}{2}\ z_{\nu+2}(y).
\phantom{aaa} \label{zexp}
\end{eqnarray}

\section{Lattice summations} \label{ApB}
\begin{figure}[]
\begin{center}
\includegraphics[clip,width=0.37\textwidth]{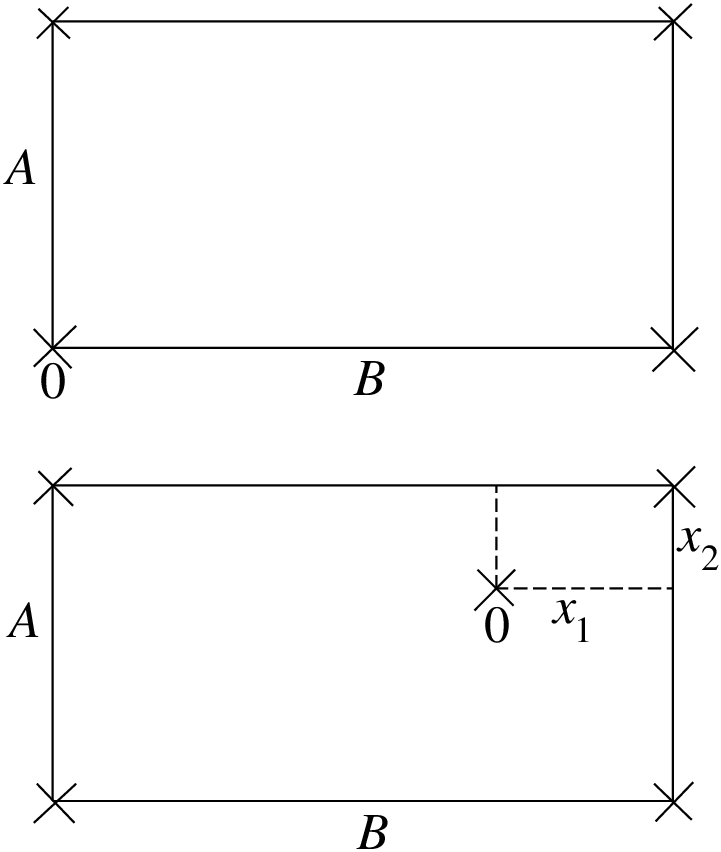}
\caption{The periodic rectangle $B\times A$ with (i) the reference site $0$
being on the rectangle; (ii) the reference site $0$ being shifted from the
rectangle.}
\label{fig:Rectangles}
\end{center}
\end{figure}

Using a sequence of transformations for Coulomb lattice sums,\cite{Samaj,TS}
we derive the interaction energy as a series of Misra functions for two
geometries presented in Fig. \ref{fig:Rectangles}.

Let the periodic rectangular lattice of sides $B\times A$ first includes 
the reference site $0$.
Taking the particles on the rectangle vertices to have charge $Q=1$,
the interaction energy of the reference particle with all others (but not
with itself) can be written as
\begin{equation}
E_1 = \sum_{j,k=-\infty\atop (j,k)\ne (0,0)}^{\infty} \frac{1}{\sqrt{(jB)^2+(kA)^2}} 
+ {\rm backgr} .
\end{equation}
This energy can be expressed as\cite{Samaj,TS} 
\begin{equation}
E_1 = \frac{1}{\sqrt{AB}} \frac{1}{\sqrt{\pi}} \Sigma_1(\psi) ,
\qquad \psi = \frac{B}{A} ,
\end{equation} 
where
\begin{widetext}
\begin{equation} \label{s1}
\Sigma_1(\psi) = 4 \sum_{j=1}^\infty \left[z_{3/2}\left(\frac{j^2}{\psi}\right)
+ z_{3/2}\left(j^2\psi\right) \right] +
8 \sum_{j,k=1}^\infty z_{3/2}\left(\frac{j^2}{\psi}+k^2\psi\right) - 4\sqrt{\pi} .
\end{equation}

If the reference site $0$ is shifted with respect to the rectangular lattice
as indicated in Fig. \ref{fig:Rectangles}, the energy 
\begin{equation}
E_2 = \sum_{j,k=-\infty}^{\infty} \frac{1}{\sqrt{(x_1+jB)^2+(x_2+kA)^2}} 
+ {\rm backgr} 
\end{equation}
is expressible as\cite{Samaj,TS} 
\begin{equation}
E_2 = \frac{1}{\sqrt{AB}} \frac{1}{\sqrt{\pi}} 
\Sigma_2(\psi,\alpha_1,\alpha_2) ,
\quad \psi = \frac{B}{A} , \quad \alpha_1 = \frac{x_1}{B} , 
\quad \alpha_2 = \frac{x_2}{A} ,
\end{equation} 
where
\begin{eqnarray} 
\Sigma_2(\psi,\alpha_1,\alpha_2) = 2\sum_{j=1}^{\infty} 
\left[\cos(2\pi j \alpha_1) z_{3/2}\left(\frac{j^2}{\psi}\right)
+\cos(2\pi j \alpha_2)z_{3/2}\left(j^2\psi\right)\right] 
\phantom{aaaaaaaaaaaaaaaaaaaaaaaaaaaaaaa} \nonumber\\ 
+ 4\sum_{j,k=1}^\infty \cos(2\pi j \alpha_1)\cos(2\pi k \alpha_2)
z_{3/2}\left(\frac{j^2}{\psi}+k^2\psi\right) + \sum_{j,k=-\infty}^\infty 
z_{3/2}\left[(j+\alpha_1)^2\psi+\frac{(k+\alpha_2)^2}{\psi}\right]
- 2\sqrt{\pi} . \label{s2}
\end{eqnarray}

From the representation (\ref{znu0y}) of the Misra functions $z_{3/2}$ 
it is simple to show that
\begin{equation}
\lim_{\alpha_1,\alpha_2\to 0} \left[
\Sigma_2(\psi,\alpha_1,\alpha_2)
- \sqrt{\frac{\pi}{\alpha_1^2\psi+\alpha_2^2/\psi}} \right]
= 2 \sqrt{\pi} + \Sigma_1(\psi) ,
\end{equation}
where the subtraction of the singular term from $\Sigma_2$ corresponds
to the interaction of the reference site $0$ with its position on the
rectangular lattice when $\alpha_1,\alpha_2\to 0$.

\section{Small-$q$ expansion of $E_{\rm I}$} \label{ApC}
We aim at calculating the small-$q$ expansion of the optimized 
energy of structure I, $E_{\rm I}(q)$, from Eq. (\ref{e1}), up to the $q^2$ term.
The absolute term can be obtained straightforwardly by setting the limiting 
$q\to 0$ value of the parameters $\Delta_1=\Delta_2=1$ $(\psi=\sqrt{3})$
in the first two summands, to get
\begin{equation} \label{e1q0}
\frac{1}{4\sqrt{\pi}} \left[ \frac{1}{2} \Sigma_1\left(\sqrt{3}\right)
+ \frac{1}{2}\Sigma_2\left(\sqrt{3},\frac{1}{2},\frac{1}{2}\right) \right] 
= \frac{2}{3} E_0(0) + \frac{1}{3} E_0(0) = E_0(0).
\end{equation}
The contributions of these first two summands to the term linear in $q$
cancel with each other, so that the $q$ term is given by the next two summands 
in (\ref{e1}), where we set $\Delta_1=1$, $d=0$ and $\psi=\sqrt{3}$,
\begin{equation} \label{e1q1}
\frac{1}{4\sqrt{\pi}} \left[ \Sigma_2\left(\sqrt{3},\frac{1}{3},0\right)
+ \frac{1}{2}\Sigma_2\left( \sqrt{3},\frac{1}{6},\frac{1}{2}\right) 
\right] q  = \left( \sqrt{3}-1\right) E_0(0) q.
\end{equation}
In order to derive the $q^2$ term, we use the Taylor expansion in $q$ of 
the internal parameters:
\begin{eqnarray} 
\Delta_1& \approx & 1+(c_1+\delta c) q + c_{21} q^2 , \nonumber\\
\Delta_2& \approx & 1+(-c_1+\delta c) q + c_{22} q^2 , \nonumber\\
d & \approx & d_1 q. \label{deld}
\end{eqnarray}
Expanding Eq. (\ref{e1}) in $q$ and using the expansion (\ref{zexp}), we get
\begin{equation} \label{e1exq2}
E_{\rm I}(q) = E_0(0) + \left( \sqrt{3}-1\right) E_0(0) q
+\left[ a_2 c_1^2 + a_1 c_1 + a_0 (\delta c)^2 + 
\frac{\sqrt{3}-3}{6} E_0(0) \right] q^2 + {\cal O}\left(q^3\right) ,
\end{equation}
where
\begin{eqnarray} 
a_1 &=& \frac{1}{4\sqrt{\pi}} \Bigg\{
-\frac{8\pi}{9}\sum_{j=1}^{\infty} j \sin\left(\frac{2\pi j}{3}\right)
z_{3/2}\left( \frac{j^2}{\sqrt{3}}\right)
-\frac{16\pi}{9} \sum_{j,k=1}^{\infty} j \sin\left( \frac{2\pi j}{3} \right)
z_{3/2}\left( \frac{j^2}{\sqrt{3}}+k^2\sqrt{3}\right) \nonumber \\
& & - \frac{4\sqrt{3}}{9} \sum_{j,k=-\infty}^{\infty} \left( j+\frac{1}{3}\right) 
z_{5/2}\left[\left( j+\frac{1}{3}\right)^2\sqrt{3}+\frac{k^2}{\sqrt{3}}\right]
-\frac{16\pi}{9} \sum_{j=1}^{\infty} j \sin\left(\frac{\pi j}{3}\right)
z_{3/2}\left( \frac{j^2}{\sqrt{3}}\right) \nonumber\\
& & - \frac{32\pi}{9} \sum_{j,k=1}^{\infty} j 
\sin\left(\frac{\pi j}{3}\right) (-1)^k
z_{3/2}\left(\frac{j^2}{\sqrt{3}}+k^2\sqrt{3}\right)
- \frac{8\sqrt{3}}{9} \sum_{j,k=-\infty}^{\infty} \left( j+\frac{1}{6}\right) 
z_{5/2}\left[\left(j+\frac{1}{6}\right)^2\sqrt{3}+\frac{(k+1/2)^2}{\sqrt{3}}
\right] \Bigg\} \nonumber\\
&\approx& -0.9822578883980, \label{a1}
\end{eqnarray}
\begin{eqnarray} \label{a2}
a_2 &=& \frac{1}{4\sqrt{\pi}} \Bigg\{
-\frac{8\pi^2}{9} \sum_{j=1}^{\infty} j^2 (-1)^j 
z_{3/2}\left(\frac{j^2}{\sqrt{3}}\right)
-\frac{16\pi^2}{9} \sum_{j,k=1}^{\infty} j^2 (-1)^j (-1)^k
z_{3/2}\left(\frac{j^2}{\sqrt{3}}+k^2\sqrt{3}\right) \nonumber\\
& & +\frac{1}{6\sqrt{3}} \sum_{j,k=-\infty}^{\infty} (2j-3) 
z_{5/2}\left[\left(j+\frac{1}{2}\right)^2\sqrt{3}
+ \frac{1}{\sqrt{3}} \left(k+\frac{1}{2}\right)^2 \right] \nonumber\\
& & +\frac{4}{3}\sum_{j,k=-\infty}^{\infty} \left(j+\frac{1}{2}\right)^2 
z_{7/2}\left[\left(j+\frac{1}{2}\right)^2\sqrt{3}+\frac{1}{\sqrt{3}}
\left( k+\frac{1}{2} \right)^2 \right] \Bigg\} \nonumber\\
& \approx & 0.62793330025966
\end{eqnarray}
and $a_0\approx 1.09206318008$ (the series expression for $a_0$ is also 
at our disposal).
Here, the terms proportional to $c_1 q$ and $(c_{21}-c_{22})q^2$ vanish since
their prefactors contain a sum from $j=-\infty$ to $\infty$ of the summands
$(j+1/2)z_{\nu}[(j+1/2)^2...]$ which cancel in pairs $\{j=0, j= -1\}$, 
$\{j=1, j=-2\}$, etc. 
The same holds for the mixed term proportional to $c_1\delta c\ q^2$
whose prefactor contains the sum of $(j+1/2)^3 z_{\nu}[(j+1/2)^2...]$.
The terms proportional to $d_1 q^2$, $\delta c\ q$, $\delta c\ q^2$ and 
$(c_{21}+c_{22})q^2$ vanish as well, though the cancellation is more tricky.
The minimization of the contribution $a_0 (\delta c)^2$ to the energy $E_I$ 
implies $\delta c=0$ because $a_0$ is positive.
We are left with the only optimization parameter $c_1$.
Taking $\partial E_{\rm I}/\partial c_1 = 0$ yields
\begin{equation} \label{c1}
c_1 = -\frac{a_1}{2a_2} \approx 0.78213552935625.
\end{equation}
Inserting the values of $a_1$, $a_2$, $c_1$ and $\delta c=0$ into 
Eq. (\ref{e1exq2}), we get with a high precision the value of 
the coefficient at $q^2$ in Eq. (\ref{e1exp}).

\end{widetext}

\end{document}